\documentclass[11pt]{article}
\usepackage{graphicx}
\begin{document}

{\scriptsize \em
15th. October 2014}

\noindent Physica B, to be published

\noindent
{\large{\bf Energy versus wave vector for the conduction band of La-doped
SrTiO$_3$ interpreted by mixed-polaron theory}}

\vskip 2em

\noindent
{\bf D.M. Eagles}

\noindent
\vskip 1em
\noindent
{\em 19 Holt Road, Harold Hill, Romford, Essex RM3 8PN, England} 

\vskip 1em
\noindent
{\em E-mail address: d.eagles@ic.ac.uk}\newline

\vskip 2em
\noindent
Short title: Dispersion of the conduction band of La-doped 
SrTiO$_3$


\vskip 2em \noindent
{\bf Abstract}

\noindent
An unusual published energy versus wave vector curve extending to the
zone edge determined by ARPES in 5\% La-doped SrTiO$_3$ is interpreted
using mixed-polaron theory.  After modification of the theory to permit
tight-binding bare energy bands and anisotropy, a fit is made to the
24 points nearest to the Fermi energy, adding the constraints (a) that the
Fermi volume is approximately equal to that determined from the carrier
concentration, and (b) that the transport mass from experiments is
approximately twice the bare mass.  In order to fit the remaining five
of 29 observed points we empirically add the effect of a kink.
The fit to all points is good.  From the parameters of the fit to
the 24 points it appears that: (a) The energy overlap integral for the
narrow band in the heavy direction has the opposite sign to that for
the wide band, implying that it is energetically favourable for the
nearly-small polaron to be centred in between two sites; and (b) The
narrower energy band in the theory may be nearly-small-polaron-like in
the heavy direction but large-polaron-like in the other two directions.
A simpler energy versus wave vector curve for a 1\% La-doped crystal is
discussed more briefly.

\noindent {\bf 1. Introduction}

\noindent

Two recent papers have provided ARPES results on SrTiO$_3$ \cite{Ch10,
Me10}.  The first of these confirms elliptic Fermi surfaces in reduced
SrTiO$_3$ with a carrier concentration stated to be $n=1.4\times
10^{20} $cm$^{-3}$, although we think that Chang et al. meant that
this was the carrier concentration per valley, and the total carrier
concentration was three times larger.   They found masses in the heavy and
light directions of 7$m_e$ and 1.2 $m_e$ at 150 K.  These masses would be
inconsistent with the observed Fermi energy of 0.046 eV if the total $n$
was as stated by them.\footnote {Dr. Chang has kindly confirmed that
the carrier concentration they mentioned was per valley.} The second
paper \cite{Me10} finds $E-k$ curves in a particular direction at 10 K for
two La-doped crystals with La fractions $x$ of 0.01 and 0.05.
The Fermi surface is reached for wave vectors well short of the zone
edge in the  $x=0.01$ crystal, but for the $x=0.05$ crystal the $E-k$
curve is shown right up to the zone edge in the $\Gamma X$-direction.
The form of the dispersion is similar to that expected from what we
call mixed-polaron theory, previously used to interpret many different
experiments on SrTiO$_3$ \cite{Ea69a, Ea84, Ea85, Ea96}.

In this type of theory the polarons are composed of mixtures of two types,
large and what we call nearly small \cite{Ea66}, and in the application of
the theory to lightly-doped SrTiO$_3$ the wave functions are composed of
roughly equal proportions of the two types.  Since the ratio of occupation
probability of the two types is sensitive to small changes of parameters
such as temperature or bare masses, properties of mixed polarons can
change strongly with changes in parameters.  Also, since some properties
such as those depending on the density of states at the Fermi energy
are more influenced by the nearly-small-polaron component of
the wave function, while other properties such as transport mass are more
influenced by the large-polaron component, in some experiments the  
polarons may seem small-polaron like and in others large-polaron like.
If the bottom of the wide band has a lower energy than the bottom of the
narrow band, then the effective mass increases as the energy increases,
because the amount of nearly-small polaron-character increases as the
energy rises.  It was this property that is visible in the dispersion
for the 5\%-doped crystal from ARPES that first made me think that
mixed-polaron theory would be relevant.

Mixed polaron theory has been used previously to interpret: 1. Large
increases in effective masses in Zr-doped SrTiO$_3$ with some Zr
replacing Ti \cite{Ea69a}, deduced \cite{Ea69b} from results on
magnetic-field penetration depths in Zr-doped SrTiO$_3$ \cite{Hu66};
2. Results on infrared absorption \cite{Ea84} in reduced SrTiO$_3$ at
two temperatures obtained by Bursian et al. \cite{Bu76}; 3. Increases in
masses with temperature \cite{Ea85} as deduced from specific-heat and
thermoelectric-power measurements \cite{Am66,Fr64,Fr66}; 4. Reductions
in plasma frequencies as temperature rises in Nb-doped SrTiO$_3$
\cite{Ea96} found by Gervais et al \cite{Ge93}.   It has also been used,
with further assumptions, to give a plausible explanation \cite{Ea71}
as to why the density-of-states effective mass in SrTiO$_3$ determined by
tunnelling \cite{Sr69,Sr70} is smaller than that found by other methods.

Recently the term mixed-polaron has been used to indicate polarons with
both coupling to breathing-modes and Su-Schrieffer-Heeger type of coupling
\cite{He13}.  We draw attention to the fact that this
recent use of the term mixed polaron has no relation to the other meaning
in our work.

Devreese  et al. \cite{De10} claim that large-polaron theory alone is
sufficient to interpret infrared data of van Mechelen et al.  \cite{Me08}.
However, because they can only fit the data for photon energies greater
than 300 meV, we do not find their arguments convincing.  Nevertheless,
one point made by Devreese et al. is that many-body effects reduce
the absorption by large polarons by quite a large fraction for carrier
concentrations of 10$^{20}$ cm$^{-3}$.   How much such effects change
the three contributions to the absorption involving the nearly-small
component of the wave functions in either initial or final states or
both in mixed-polaron theory has not been studied yet, but such effects
may have to be taken into account in further study of optical properties
in mixed-polaron theory for higher carrier concentrations than those of
1.1$\times 10^{18}$ cm$^{-3}$ examined by Bursian et al..

In this paper we show that mixed-polaron theory can be used to interpret
the $E-k$ curves right up to the zone edge in 5\% La-doped SrTiO$_{3}$
obtained by Meevasana et al. \cite{Me10}, but the energy overlap integral
for the narrow band in the heavy direction is probably of opposite sign
to that of the wide band.  This difference in sign from that found in the
lightly-doped samples is tentatively attributed to bond-centred nearly
small polarons having the lowest energy above some critical doping.
In 1\% La-doped SrTiO$_3$, there is too restricted a range of wave
vectors for which observations have been recorded to permit confident
determination of all parameters in a multiparameter fit, but we have
found two sets of parameters which gives fairly good fits to the data.

In section 2 we write the basic equations of mixed-polaron theory modified
from a previous formulation \cite{Ea85} by use of a tight-binding model
with anisotropy for the bare conduction band.   We also modify the theory
for energies more than about 0.02 eV below the Fermi energy, where 0.02
eV is the approximate energy of a low energy phonon causing a kink in
the E(k) curve at its energy.   In section 3 we discuss how we fit the
data, and give results for the data fitting for the 5\%-doped crystal,
and a briefer discussion of the results for the 1\%-doped crystal.
In section 4 we show how to calculate the transport mass approximately
for the 5\%-doped crystal in our model.  In section 5 we show the
calculated shape of the Fermi surface for the 5\%-doped crystal, and
section 6 gives a qualitative discussion of whether site- or bond-centred
nearly small polarons have the lower energy.  A brief discussion of the
plausibility of the values of parameters found is given in section 7.

\noindent {\bf 2. Mixed-polaron theory for SrTiO$_3$}

\noindent
We use mixed-polaron theory in the formulation of \cite{Ea85}, except
that we first modify this theory slightly to use tight-binding  forms
for both the wide and narrow bands, and to allow different dispersions in
different directions.  Meevasana et al. \cite{Me10} find the dispersion
in the direction parallel to the surface, labelled the $x$-direction,
for a small value of the reduced (moduli a reciprocal lattice vector) wave
vector in the direction perpendicular to the surface ($z$-direction), and
we take into account dispersion in both these directions.  We assume
that, when the mixed-polaron energy is within a low energy phonon
$\hbar \omega \approx 0.02$ eV  
of the Fermi energy,
the energies $E_w$ and $E_n$ of the wide and narrow bands at wave
vector components $k_x,k_z$ in the two directions measured from the
energy of the narrow band at ${\bf k}=0$ are given by
\begin{equation}
E_w(k_x,k_z)=2t_{wh}[1- {\rm cos}(k_xa)]+2t_{wl}[1-{\rm cos}(k_za)]
-\delta \hbar \omega,
\end{equation}
\begin{equation}
E_n(k_x,k_z)=2t_{nh}[1- {\rm cos}(k_xa)]+2t_{nl}[1-{\rm cos}(k_za)],
\end{equation}
where $\delta \hbar \omega$ is the energy difference of the narrow band
and wide band at $k=0$, and $a$ is the lattice constant.  We ignore
any slight orthorhombicity at low temperatures.  Since the dispersion
extends to the zone edge in the $x$-direction, we assume that this is
the heavy-mass direction, and the suffices $h$ and $l$ above denote
"heavy" and "light".

The restriction of use of the energies of equations (1) and (2) to a
region close to the Fermi level is  based on the fact that Meevasana
et al. \cite{Me10} find that an interaction corresponding to a coupling
$\lambda=0.3$ occurs with phonons of energy $\hbar\omega_0 \approx 0.02$
eV in both crystals of SrTiO$_3$ studied, although more clearly shown
for the 1\%-doped crystal.   This interaction gives rise to a change
of slope of the $E(k)$ curve and a sudden drop in the energy at the
energy of the phonon.  An interaction of strength $\lambda$ implies
that the wide-band energy overlap integrals $t_{wl}$ and $t_{wh}$
will increase by factors $(t_{bl}/t_{wl})/(t_{bl}/t_{wl}-\lambda)$
and $(t_{bh}/t_{wh})/(t_{bh}/ t_{wh}-\lambda)$  at energies more than
the relevant phonon energy below the Fermi energy, where $t_{bl}$ and
$t_{bh}$ are the bare overlap integrals.  Thus, when the mixed-polaron
energy is more than the appropriate phonon energy $\hbar\omega_0$ below
the Fermi energy we replace equation (1) by
\begin{equation}
E_w(k_x,k_z)=2t_{wh2}[1- {\rm cos}(k_xa)]+2t_{wl2}[1-{\rm cos}(k_za)]
-\delta \hbar \omega-d\hbar\omega
\end{equation}
where
\begin{equation}
t_{wh2}=t_{wh}\{(t_{bh}/t_{wh})/[(t_{bh}/t_{wh})-\lambda]\},\;\;
t_{wl2}=t_{wl}\{(t_{bl}/t_{wl})/[(t_{bl}/t_{wl})-\lambda]\},
\end{equation}
and $d$
is introduced to make the mixed-polaron energy drop
suddenly by $\lambda\hbar\omega_0$  at a critical wave number $k_{xc}$.
An explicit expression for $d$ will be given shortly.

We ignore changes in the narrow band when the total polaron energy is
more than the relevant phonon energy below the Fermi energy.

Introducing notations similar to those of \cite{Ea85},
we write 
\begin{equation} 
(t_{wh}+t_{nh})/\hbar \omega=s_h,
\;(t_{wh}-t_{nh})/\hbar\omega=d_h, 
\end{equation} 
\begin{equation}
(t_{wl}+t_{nl})/\hbar \omega=s_l, \;(t_{wl}-t_{nl})/\hbar\omega=d_l,
\end{equation} 
\begin{equation} (t_{wh2}+t_{nh})/\hbar
\omega=s_{h2}, \;(t_{wh2}-t_{nh})/\hbar\omega=d_{h2},
\end{equation} 
\begin{equation} 
(t_{wl2}+t_{nl})/\hbar \omega=s_{l2},
\;(t_{wl2}-t_{nl})/\hbar\omega=d_{l2}.  
\end{equation} 
Then we find that
the state of the lowest mixed-polaron band with wave vector $k_x,k_z$ in
the $xz$-plane has an energy $E_g(k_x,k_z)$ measured from the Fermi energy
given by 
\begin{equation} E_{g}(k_x,k_z)/\hbar \omega = z, 
\end{equation}
where $\hbar\omega=0.09$ eV is a weighted mean longitudinal polar
optical phonon energy, and, if  $|k_x|>|k_{xc}|$, where $k_{xc}$ is the 
wave number where the kink starts, 
\begin{equation}
z=s_h[1-{\rm cos}(k_xa)]+s_l[1-{\rm
cos}(k_za)] -\delta/2-y/2+[E_n(0)-E_F]/\hbar \omega.  
\end{equation} 
Here
\begin{equation} 
y=(b^2+x^2)^{1/2}, 
\end{equation} 
where 
\begin{equation}
b=2|U|/\hbar \omega 
\end{equation} 
is twice the magnitude in phonon
units of the matrix element of the Hamiltonian between orthogonalised
large and nearly-small polarons of the same wave vector (assumed to
be independent of wave vector), $a=0.3905$ nm, and 
\begin{equation}
x=2d_h[1-{\rm cos}(k_xa)]+2d_l[1-{\rm
cos}(k_za)]-\delta.  
\end{equation}

By the requirement that the drop in energy at $k_{xc}$ is
$\lambda\hbar\omega_0$, we find the
following equation for $d$ of equation (3):  
\begin{equation}
j+0.5d+0.5[b^2+(g-d)^2]^{\frac{1}{2}}=\lambda\omega_0/\omega,
\end{equation}
where
\begin{eqnarray}
j=(s_h-s_{h2})[1-{\rm{cos}}(ak_{xc})]+(s_l-s_{l2})[1-{\rm{cos}}(ak_z)]
\nonumber\\
-0.5\{b^2+[2d_h(1-{\rm cos}\overline{ak_{xc}})+2d_l((1-{\rm cos}
\overline{ak_z})-\delta)]^2\}^{\frac{1}{2}}
\end{eqnarray}
and 
\begin{equation}
g=2d_{h2}[1-{\rm cos}(k_xa)]+2d_{l2}
[1-{\rm cos}(k_za)]-\delta.
\end{equation}

Equation (15) reduces to
\begin{equation} 
d=[2(j-\lambda\omega_0/\omega)^2-0.5(b^2+g^2))/(2\lambda\omega_0/\omega
-g-2j).
\end{equation}

From numerical data given to me by Dr. Meevasana, it looks as if the first
threshold energy $\hbar \omega_0$ is closer to 0.018 eV than 0.02 eV.
Also from the data it appears that $k_{xc1}\approx 3.5$ nm$^{-1}$ for the
5\%-doped crystal and $\approx 2.65$ nm$^{-1}$ for the 1\%-doped crystal.
A first guess at the second threshold energy would be the energy of the
second longitudinal optical phonon ($\hbar\omega_2\approx$ 0.058 eV 
\cite{Ea65,Ba66}).  However, data of Meevasana et al \cite{Me10}, appear
to show that this second threshold is somewhere between 0.04 and 0.05 eV.
For energies more than $\hbar\omega_2$ below the Fermi energy, we could
introduce further changes, but we prefer to deal with the few observed
points below this energy in a different way, as discussed in the next
section.

\noindent
{\bf 3. Method of fitting data} 

\noindent
{\em 3.1 5\%-doped crystal}

\noindent
In this section we describe how we fit the data of figure 3(b) of
Meevasana et al \cite{Me10}.  We concentrate first on fitting the 24
points nearest to the Fermi energy.  The remaining five of 29 observed
points are less accurate, but we can also fit them if we  include a kink
\cite{Da03,Cu04} associated with the second lowest longitudinal
optical phonon, with the kink determined by two parameters related
to its position and slope.

Meevasana (private communication) estimates that, for a 24 eV photon
energy, the reduced wave vector $k_z$ (modulo a reciprocal lattice vector)
is about $k_z=-1$ nm$^{-1}$.  However, the quantities appearing in this
estimate may not be known accurately, and so we shall treat $|k_z|$ as
adjustable initially, but check that the best-fit value is not necessarily
inconsistent with Meevasana's estimate.  Then we have eight parameters
appearing in the theory, i.e. $t_{wh}$, $t_{nh}$, $t_{wl}$, $t_{nl}$,
$b$, $\delta$, $[E_n(0)-E_F]$, and $|k_z|$.  We estimate  $t_{wh}$ in terms of
$t_{wl}$ using the assumption that that the anisotropy of weak-coupling
polaron masses is the same as that of bare masses.  This is approximately
true for weak and intermediate coupling in the Holstein model \cite{Al08},
but is poorer when long-range forces are included \cite{Ka68}.

Thus we have seven parameters to fit the first 24 points.  We also impose
constraints (a) that the Fermi volume is in agreement with that calculated
from the known carrier concentration, and (b) that the transport mass
as determined by experiments \cite{Ma11} is approximately twice the
bare mass, and (c) a constraint on $t_{wl}$ based on estimates 
of lower and upper limits of by what factor it is reduced from its bare value
$t_{wb}$.
From figure 3 of \cite{Ma11} we estimate very approximately that the
average bare mass of the two light bands up to the highest energy 110
meV above the bottom of the heavy-mass band shown in the figure is about
0.54 $m_e$, and that of the heavy mass band is about 7.3 $m_e$, both with
appreciable uncertainties due to approximate methods of
making the estimates.  According to \cite{Ma11} there is about a factor
of two increase in transport masses over bare masses at high carrier
concentrations.  In our theory this is partly due to the admixture of
narrow-band states to the mixed polarons, and partly due to increases
in the wide-band masses due to electron-phonon interactions.  

Since the Fermi wave vector reaches the zone edge in the heavy direction,
the volume of the Fermi surface determined by the carrier concentration
requires that in the light direction the Fermi wave number is much
smaller.  Besides the data points we fit, we also require that the Fermi
volume is consistent with that expected from the known doping level.
Then, with an effective doping level of 5.6\% \cite{Me10}, the carrier
concentration $n_v$ per valley is $n_v\approx 3.13 \times 10^{20}$
cm$^{-3}$, and so the Fermi volume per valley is $4\pi^3\times 3.13\times
10^{20}$ cm$^{-3}$.  Hence we have the constraint
\begin{equation}
(1/2)\int_{-\pi/a}^{\pi/a} dk_x \int_0^{2\pi} d\theta\; k_{Fp}^2=
4\pi^3\times 3.13\times 10^{20} {\rm cm}^{-3}.  
\end{equation}
Here $k_{Fp}$ denotes the magnitude of the $yz$-plane component of the
Fermi wave vector for a given $k_x$.   For the weighting we chose for this
constraint, the value of the Fermi volume for the best fit was accurately
in agreement with the value based on the estimated carrier concentration.
We made approximations for the double integral. 

Our constraint on $t_{wl}$ is estimated as follows. The value of the
bare overlap integral $t_{wb}$ in the light direction which gives a
bare mass of 0.54 $m_e$ is $t_{wb}=463$ meV.  The overlap $t_{wl}$
will be reduced from the bare overlap integral by a factor which is
almost certainly between a factor of (1/1.3)=0.77 assuming that the
only reduction due to phonons is by the mode with energy of about 0.02 eV
with a coupling paremeter of 0.3 as found empirically in \cite{Me10},
and by a factor which also includes the effect of unscreened interactions
with longitudinal optical modes as estimated by use of \cite{Ea65} plus
effects of anisotropy on weak-coupling polarons by Kahn \cite{Ka68}.
With the estimate of anisotropy of bare masses given below this gives us a
lower limit on $t_{wl}/t{wb}$ approximately equal to [1/(1.3+0.45)]=0.57,
where the quantity 0.45 comes from $(2.3/6)(0.54^{\frac{1}{2}})\times
1.6$, where 1.6 is taken approximately from figure 2 of \cite{Ka68} with
a bare-mass anisotropy of 7.3/0.54=13.5, 2.3 is the coupling constant
$\alpha$ for $m=m_e$ in SrTiO$_3$ \cite{Ea65}, and the mass increase in
weak-coupling polaron theory is by a factor $(1+\alpha/6)$ \cite{Fr54}.
Thus we estimate that 264 meV $< t_{wl} < 357$ meV, and we take these
as 95\% confidence limits.

How to impose the constraint involving the transport mass will be discussed
in section 4.

The seven parameters determined by fitting the data with constraints as
described above have the values $t_{nh} = -7.69$ meV,
$t_{nl} = 170$ meV, $b = 0.704$ , $\delta
= 0.741$, $[E_n(0)-E_F] = 26.8$ meV,
$k_z=0.462$ nm $^{-1}$, $t_{wl}=317$ meV,
and the deduced value of the other parameter $t_{wh}=$
23.5 meV.  

The results for the parameters determined by the best fit are shown in
figure 1.  The other five unfitted points are also shown on the figure,
both without and with a conjectured kink effect determined by
two extra parameters for the 5\%-doped crystal.  The root-mean-square
difference between the observed and calculated value of energy
for the 24 fitted points is 0.72 meV.

The calculation of the Fermi volume used in the fit does
not take into account any possible effects due to kinks.   However,
the kinks are expected to have the same depths in all directions,
since these depths depend on the contribution of interactions with the
relevant phonon modes to the electronic energy, and so the calculated
Fermi volume should not be influenced by kinks.

The value of the ratio of the overlap integral $t_{wb}$ associated with
the bare mass of about 0.54 $m_e$ in the light direction determined
approximately from figure 2 of \cite{Ma11} satisfies $t_{wb}/t_{wl}
\approx$ 1.46.  The transport mass at high carrier concentrations is
about twice the bare mass \cite{Ma11}.  Thus the above ratio implies
that 46\% of the increase in the transport mass is due to increases of
the wide-band mass above the bare mass, with the remainder due to the
admixture of narrow-band states in the mixed-polaron wave function.

We note that $t_{nh}$ is probably negative, i.e. a negative mass at $k=0$
for the nearly-small polaron band.  This could be expected if the energy
of the system comprised of bond-centred nearly small polarons is lower
than for the case of site-centred polarons.  For further discussion of
this see section 6.  The value of $t_{nl}$ determined by our fitting
is such that $t_{nl}/t_{bl}=0.37$.  This is higher than might be
expected since, for site-centred polarons in lightly-doped crystals,
the corresponding ratio is of the order of 0.01 \cite{Ea85}.  For the
heavy direction, the most probable ratio $|t_{nh}/t_{bh}|\approx 0.22$,
which is also larger than the ratio of average overlaps
of the order of 0.01 in lightly-doped crystals.  We suggest tentatively
that the narrow band may be composed of polarons which are nearly small
and bond centred in one direction, but large in the other two directions,
and that the larger ratio of narrow-band to bare overlap integrals may
have different causes in different directions, i.e. because the polarons
are large-polaron like in the two light directions but, for the heavy direction
because of the smaller coupling to phonons for bond-centred nearly small
polarons because they are more spread out.  The problem of whether it is
plausible to have two types of states with a fairly small matrix element
of the Hamiltonian between them but with masses not drastically different
in two directions might require quite a lot of work to resolve.

The value of $|k_z|\approx 0.46 $ nm$^{ -1}$ determined by our fit is
slightly smaller in magnitude than the value of the order of 1
nm$^{-1}$ estimated by Dr. Meevasana in calculations accompanying a
letter, but the uncertainties in quantities put into these calculations
make it look as if his estimated value would be consistent with our 
value.   Also Dr. Meevasana in his letter states that the observed
dispersion in the $\Gamma-X$ direction is approximately the same as if
$|k_z|$ were zero.

The second kink in figure 1 commences at an energy between about 40 and 48
meV below the Fermi energy.  This is lower than the energy of the middle
energy longitudinal optical phonon of 58 meV in lightly-doped samples
\cite{Ea65,Ba66}. Several possible explanations for this difference
are: (a) the phonon energies may be lower in heavily doped crystals,
(b) dispersion of the mode in a downward direction  may occur in some
directions in some models \cite{Co64}, or (c) some coupling to the
transverse mode may occur when one gets away from low wave vectors.
The middle transverse mode at ${\bf k}=0$ has an energy of 22 meV in
lightly-doped crystals \cite{Sp62}.

From figure 1, we can see that the reduction in energy due to interaction
with the mode involved in the second kink is somewhere between 21 and
30  meV.  The coupling constant with the middle longitudinal mode in
lightly doped crystals is $0.50(m_b/m_e)^{\frac{1}{2}}$ \cite{Ea65}.
Extrapolating from figure 2 of \cite{Ka68}, with bare transverse
and longitudinal masses of 0.54 $m_e$ and 7.3 $m_e$, we deduce
that the contribution to the polaron binding energy would be about
$0.5\times(0.54)^{\frac{1}{2}}\times 1.3 \times 58 = 28$ meV in Meevasana
et al.'s crystal if there were no reduction in coupling due to screening.
Screening may decrease this contribution to a value closer to the lower limit
of 21 meV estimated above. 

For more heavily doped crystals, another kink would be expected
starting at an energy below $E_F$ close to the highest energy longitudinal
mode of energy 99 meV of the lightly-doped crystals.

Although an expression for $b$ in our theory is given in \cite{Ea69a},
it involves a sum of terms of differing signs which come close to
cancelling, and so analysing the values of $b$ obtained here 
is not likely to be useful.  Also $\delta$ is strongly correlated with
other parameters, and so its value is not easy to interpret.

\noindent
{\em 3.2 Remarks on the 1\% doped crystal}

\noindent
Although there are 32 observed points on the $E(k)$ curve for the 1\%-
doped crystal, the range of wave vectors involved and the relative
simplicity of the curve makes it difficult to know whether any minimum
found in our non-linear least squares fitting routine is an absolute
minimum.  As for the other crystal, we first added a constraint on
$t_{wl}$,  with an assumed most probable value of 289 meV, slightly
smaller than obtained by our fit for the 5\%-doped crystal because of
smaller screening of electron-phonon interactions.  We also imposed
a constraint on $t_{nl}$ based on the value found for this for the
5\%-doped crystal, although we also tried putting both $t_{nh}$ and
$t_{nl}$ equal to 1.4 meV, a value expected if there are site-centred
polarons \cite{Ea85}.  The method of estimating $t_{nl}$ in \cite{Ea85}
is appropriate when the energy of the saddle point between two
minima in the polaron potential energy curve is smaller than the phonon
energy $\hbar\omega.$ For the 1\%-doped crystal we took the bare mass in
the heavy direction to be 5.6 $m_e$ rather than 7.3 $m_e$ in the 5\%-doped
crystal because band-structure calculations indicate that the bare mass
in this direction decreases as carrier concentration decreases.  We did
not use the constraint that the transport mass is approximately equal
to twice the bare mass for this crystal, as we did not know in advance
an approximate shape of the Fermi surface. 

For the first type of fit, with a constraint on $t_{nl}$ to be close
the value of  170 meV found from our fit to the data on the 5\%-doped
crystal, we found the following values of parameters:  $t_{nh}=-5.1$
meV, $t_{nl}=172$ meV, $b=0.716$, $\delta=0.764$, $[E_n(0)-E_F]=43.8$ meV,
$|k_z|=0.026$ nm$^{-1}$, $t_{wl}=315$ meV, with the deduced value of
$t_{wh}=$30.3 meV.   The root-mean-square difference between theory and
experiment for the 32 points was 1.18 meV.

With the second procedure, with $t_{nh}$ and $t_{nl}$ 
equal to 1.4 meV, as expected for site-centred nearly small
polarons \cite{Ea85}, we found $b=0.681$, $\delta=0.760$, $[E_n(0)-E_F]=
41.8$ meV, $|k_z|=0.0$ nm$^{-1}$, $t_{wl}=297$ meV, with the deduced value
of $t_{wh}=28.7$ meV.  The root-mean-square difference between theory and
experiment for the 32 points for this case is 1.39 meV, slightly larger
than for the first type of fit, but with more constraints on parameters.
So we cannot say definitely whether site or bond-centred polarons are
preferred in the 1\%-doped crystal.

We show the fit obtained with the first set of parameters in figure 1.  

\noindent
{\bf 4. Approximate calculation of the transport mass for the 5\%-doped crystal}

In this section we calculate an approximate value of the average over
the Fermi surface of the transport mass (reciprocal of the average
inverse mass), for the 5\%-doped crystal,  made with the approximation
that the Fermi surface is replaced by a cylinder of radius 0.87 nm$^{-1}$,
chosen to give approximately the correct Fermi volume.

At a given wave vector $\bf k$, the reciprocal of the mass in the light
direction is proportional to $t_{wl} p_{w{\bf k}}+t_{nl}[1-p_{w{\bf k}}]$, 
where
\cite{Ea69a} $p_{w\bf k}$, the fraction of the wide-band  state
in the mixed polaron of wave vector ${\bf k}$, is given by
\begin{equation}
p_{w\bf k}=(1/2)[1-r_{\bf k}/(r_{\bf k}^2+1)^{\frac{1}{2}}],
\end{equation}
with
\begin{equation}
r_{\bf k}=(E_{w\bf k}-E_{n\bf k)}/|b\hbar\omega|.
\end{equation}
Here $E_{w\bf k}$ and $E_{n\bf k}$ are given by equations (1) and
(2).  A similar expression occurs for the corresponding probability of
occupation of the narrow band for the mixed polaron of wave vector ${\bf
k}$, with the minus sign in the square bracket replaced by a plus.

The average of $p_{w\bf k}$ over the Fermi surface is given by
\begin{equation} 
Av[p_{w\bf k}]=(a/2\pi)\int_{-\pi}^\pi dk_x p_w(k_x,k_{Fz}),
\end{equation} 
where $k_{Fz}$ = 0.87 nm$^{-1}$ is the Fermi wave number
in the $z$-direction which gives the correct Fermi surface volume, with a
similar equation for the average of the probability of occupation of the
narrow-band states.  We deduce that the transport mass $m_t$ is given by
\begin{equation}
m_t \approx (3/2)\{(\hbar^2/2t_{wl}) Av[p_{w\bf k}]+(\hbar^2/2t_{nl})
[Av[p_{n\bf k}]\},
\end{equation} 
where the averages are over the Fermi surface, and  the factor 3/2
comes from only having only two out of three directions associated with
light masses.  The masses in the heavy direction do not appear in the
transport mass because the Fermi surface reaches the zone edge in the
heavy direction.   Making the calculated mass given by equation (22)
equal to the observed transport mass, which is approximately equal to
$2m_b$, where $m_b$ is the bare mass \cite{Ma11}, gives a constraint on
parameter values which was used in our data fitting for the 5\%-doped
crystal.

\noindent
{\bf 5. Shapes of Fermi surfaces}

\noindent
Although the Fermi surfaces are close to being cylindrically symmetric in
our model, the diameters of the cylinders vary with $k_x$.  In figure 2
we show the cross sections in the $xz$-plane for the parameters we found.

We see that, for the 5\%-doped crystal, the wave number in the
$z$-direction is maximised with a value of 1.10 nm$^{-1}$ at $k_x=0$,
reaches a minimum of 0.74 nm$^{-1}$ at $k_x\approx 0.66 \pi/a$, and has
a subsiduary maximum at the zone edge.  Testing whether such a shape
is valid might be a worthwhile study for future ARPES measurements.
The existence of a minimum in $k_{Fz}$ is a simple consequence of the
maximum in energy as a function of $k_x$, but the details of the curve
of $k_{Fz}$ as a function of $k_x$ depend on the model used.
The magnitude of the Fermi wave number in our model changes by less 
than 0.2\% as the azimuthal angle in the plane of the light directions
changes.

For the 1\%-doped crystal, the Fermi surface  cross section differs
slightly from an ellipse.  The value of $k_x a/\pi$ at which $k_{Fz}$
reaches (0.75)$^{\frac{1}{2}}$ of its maximum value is 0.205, which,
for an ellipse would imply $k_{Fx} a/\pi$=0.41, whereas our model gives
$k_{Fx} a/\pi$=0.47.  Thus we have something more like an ellipse which is
elongated at its ends.  The magnitude of the Fermi wave vector is 4.09
times larger in the $x$-direction than the $z$-direction.   The smaller
ratio which looks as if it occurs from the figure arises because of
our use of different scales for the two axes.

There appears to be a discrepancy between the ratios of magnitudes of axes
of the Fermi surfaces between \cite{Ch10} and \cite{Me10}.  The value of
$k_{Fx} a/\pi$=0.47 which we found above is close to what can be inferred directly
from figure 2(a) of \cite{Me10}, and so does not depend strongly on
our theory.  The Fermi surface reported in \cite{Ch10} is considerably less
anisotropic than that for the 1\%-doped crystal in \cite{Me10} despite
the probably larger carrier concentration in \cite{Ch10}.  We do not know
the cause of this discrepancy.

{\bf 6. Qualitative discussion of whether site-centred or bond-centred
nearly small polarons have lower energy}

\noindent
In order to find out whether site-centred or bond-centred nearly small
polarons are present, we would need a calculation of the total energy
of the system for both types of polaron.  For bond-centred polarons we
could use our empirical parameters to estimate this energy, but for
site-centred polarons we would need to rely on calculations.  We are
not confident that such calculations would be sufficiently accurate
to determine the sign of the probably small difference in energy of
the system for the two types.  However, we note that the ratios of the
narrow-band to bare bandwidths for what we think are bond-centred polarons
are relatively large (0.37 in the light direction in the 5\%-doped crystal)
compared with ratio of the order of
0.01 for site-centred polarons \cite{Ea85}.  As mentioned before, we
tentatively suggest that the narrow-band states we find are connected
with nearly small polarons in one direction but with large polarons in
the other two directions.

Narrow-band states start to be occupied significantly when the bottom
of the narrow band is within an energy of the order of $b\hbar\omega$
of the Fermi energy, and hence the effect of the larger width of the
narrow band can lower the total energy for Meevasana et al.'s crystal,
but not for samples with much lower carrier concentrations.   Against
this we would have to take into account the probable smaller energy
lowering of the local energy for bond- than for site-centred polarons.
This difference will be less dependent on carrier concentration than
the effect of different widths of the narrow bands because it depends
on small changes in screening.  Thus, while we cannot confirm that the
total energy is lower for bond-centred polarons, we have an argument as
to why bond-centred polarons are likely to have a lower relative energy
for large carrier concentrations, and hence that a transition betwen
types may be expected as  the carrier concentration changes.

\noindent
{\bf 7. Discussion of values of some parameters}

\noindent
In this section we discuss to what extent values found for some of the
parameters of our model are plausible.   We have already given suggestions
as to why $t_{nh}$ is negative  and its modulus rather larger than we
might have guessed (bond-centred nearly small polarons), and why $t_{nl}$
is not very small (polarons large-polaron like in two directions).
The quantities $b$ and $\delta$ are difficult to estimate, $b$ because
it is composed of several terms of different signs which come close to
cancellation \cite{Ea69a}, and $\delta$ because its value is dependent
on several other parameters.  We do not have a theory for the size of
$[E_n(0)-E_F]$, but a first guess would be that this should be larger if
$E_F$ is smaller, which agrees with our fitted values.  However,
against the reduction of $E_F$ we have a probable stronger binding of
nearly small polarons for smaller $E_F$ because of reduced screening.
We have discussed previously that our values for $|k_z|$ are probably
consistent with approximate estimates by Dr. Meevasana.

\noindent
{\bf 8. Conclusions}

\noindent
A good fit to $E-k$ curves in SrTiO$_3$ determined by Meevasana et al.
by ARPES measurements on a 5\% La-doped SrTiO$_3$ crystal is obtained
by use of mixed-polaron theory.  This type of theory has previously
been used to interpret many types of data on SrTiO$_3$.  Of particular
interest is the probable opposite sign of the energy overlap integral
associated with the narrow band in the heavy direction compared with
that for the wide band, and also the larger magnitude of the narrow-band
energy-overlap integrals than found previously for lightly-doped samples.
Both of these properties are conjectured to arise because bond-centred
nearly small polarons have lower energies than site-centred ones, and,
in the case of the second property, perhaps because the narrow-band
states found are associated with polarons that are nearly small in
one direction but large in the other two directions. For the 1\% doped
crystal we also obtained a fairly good fit to the data, but, because of the
narrow range of wave vectors involved and the simplicity of the observed
curves, we are less confident in this case that the minimum we found in
our least-squares fitting routine is an absolute minimum.

\noindent {\bf Acknowledgments}

\noindent
I should like to thank  Dr. W. Meevasana for numerical data from their
ARPES work, and for a letter showing me a calculation estimating $k_z$
for some of their measurements, and Professor J. T. Devreese for
correspondence and for a preprint of \cite{De10}.

\newpage
\begin{figure}
{\centerline{\includegraphics[width=12.6cm,]{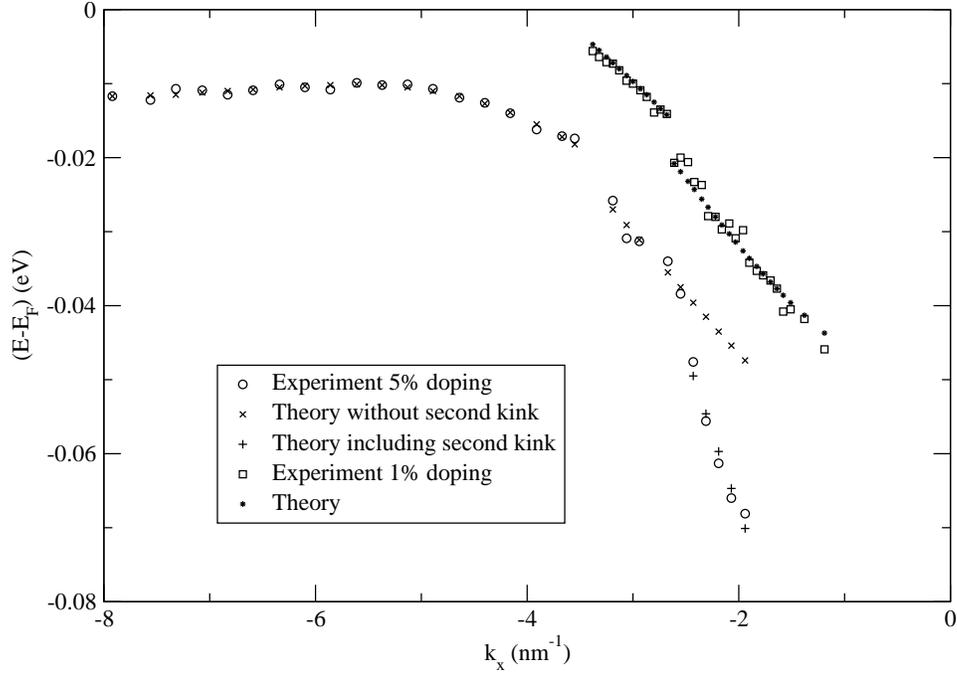}}}
\caption
{Fits with model to ARPES data from \cite{Me10} for 5\% and 1\% 
La-doped SrTiO$_3$} 
\end{figure}

\newpage
\begin{figure}
{\centerline{\includegraphics[width=12.6cm,]{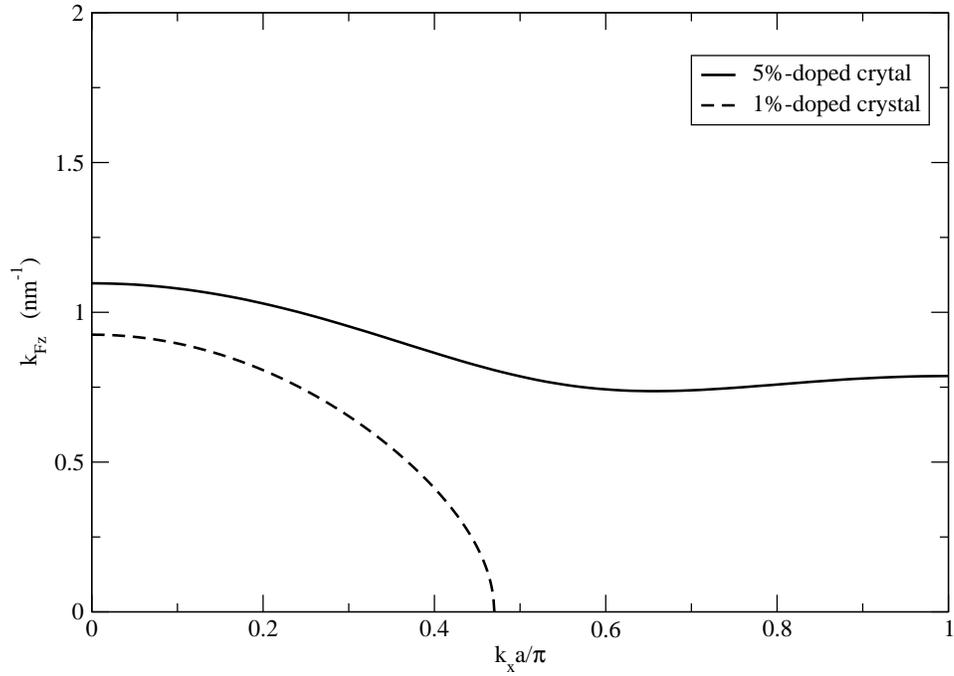}}}
\caption
{Fermi wave vector $k_{Fz}$ for both crystals in the $z$-direction
as a function of $k_xa$ according to model}
\end{figure}


\noindent
\begin{thebibliography}{99}

\bibitem{Ch10} Y.J. Chang, A. Bostwick, Y.S. Kim, E. Rotenberg,
Phys. Rev. B 81 (2010) 235109. 

\bibitem{Me10} W. Meevasana, X.J. Zhou, B. Moritz, C.-C. Chen,
R.H. He, S.-I. Fujimori, D.H. Lu , S.-K. Mo, R.G. Moore, F. Baumberger,
T.P. Devereaux,  D. van der Marel, N. Nagaosa, J. Zaanen, Z.-X. Shen,
New Journal of Physics 12 (2010) 023004.

\bibitem{Ea69a} D.M. Eagles, Phys.Rev. 181 (1969) 1278.

\bibitem{Ea84} D.M. Eagles, P. Lalousis, J.Phys.C 17 (1984) 655 
     
\bibitem{Ea85} D.M. Eagles, Physics of Disordered Materials, 
eds. D Adler, H Fritzsche and S R Ovshinsky, Plenum, New York,
1985, pp 357-367.

\bibitem{Ea96} D.M. Eagles, M. Georgiev, P.C. Petrova, 
Phys.Rev. B 54 (1996) 22.

\bibitem{Ea66} D.M. Eagles, Phys.Rev 145 (1966) 645.

\bibitem{Ea69b} D.M. Eagles, Phys.Rev 178 (1969) 668.

\bibitem{Hu66} J.K. Hulm, C.K. Jones, R.C. Miller, T.Y. Tien,
Proceedings of the Tenth International Conference on Low-Temperature
Physics Viniti, Moscow: VINITI 1966, pp 86-114.

\bibitem{Bu76} \'E.V. Bursian, G.A. Girshberg, A.V. Ruzhnikov, Fiz.
Tverd. Tela 18 (1976) 578 [Sov. Phys. Solid. State 18 (1976) 335].

\bibitem{Am66} E. Ambler, J.H. Colwell, W.R. Hosler, J.R. Schooley, 
Phys. Rev. 148 (1966) 280.

\bibitem{Fr64} H.P.R. Frederikse, W.R. Thurber, W.R. Hosler,
Phys. Rev. 134 (1964) A442.

\bibitem{Fr66} H.P.R. Frederikse, W.R. Thurber, W.R. Hosler, 
J. Phys. Soc. Jpn. Suppl. 21 (1966) 32.

\bibitem{Ge93} F. Gervais, J.-F. Servoin, A. Baratoff, J.G. Bednorz, G. Binnig,
Phys. Rev. B 47 (1993) 8187.

\bibitem{Ea71} D.M. Eagles, Phys. Stat. Sol. (b) 48
(1971) 407.

\bibitem{Sr69} Z. Sroubek, Solid State Commun. 7 (1969) 1561.

\bibitem{Sr70} Z. Sroubek, Phys. Rev. B 2 (1970) 3170.

\bibitem{He13} F. Herrera, K. Madison, R.V. Krems, M. Berciu,
Phys. Rev. Lett. 110 (2013) 223002.

\bibitem{De10} J.T. Devreese, S.N. Klimin, J.L.M. van Mechelen, D. van der
Marel, Phys. Rev. B 81 (2010) 125119.

\bibitem{Me08} J.L.M. van Mechelen, D. van der Marel, C. Grimaldi,
A.B. Kuzmenko, N.P. Armitage, H. Reyren, H. Hagerman, I.I. Mazin, 
Phys. Rev. Lett 100 (2008) 226403.

\bibitem{Ea65} D.M. Eagles, J.Phys. Chem. Solids 26 (1965) 672.

\bibitem{Ba66} A.S. Barker, Phys. Rev. 145 (1966) 391.

\bibitem{Da03} A. Damascelli, Z. Hussain, Z.-X. Shen, Rev. Mod. Phys.
75 (2003) 473.

\bibitem{Cu04} T. Cuk, F. Baumberger, D.H. Lu, N. Ingle, X.J. Zhou,
H. Eisaki, N. Kaneko, Z. Hussain, T.P. Devereaux, N. Nagaosa, Z.-X. Shen,
Phys. Rev. Lett. 93 (2004) 117003.

\bibitem{Al08} A. Alvermann, H. Fehske, S.A. Trugman, Phys. Rev. B
78 (2008) 165106.

\bibitem{Ka68} A.H. Kahn, Phys. Rev. 172 (1968) 813.

\bibitem{Ma11} D. van der Marel, J.L.M. van Mechelen, I.I. Mazin, 
Phys. Rev. B 24 (2011) 205111.

\bibitem{Fr54} H. Fr\"ohlich, Advanc. Phys. 3 (1954) 325.

\bibitem{Co64} R.A. Cowley, Phys, Rev. 134 (1964) A981.

\bibitem{Sp62} W.G. Spitzer, R.C. Miller, D.A. Kleinman, L.E. Howarth, 
Phys. Rev. 126 (1962) 1710.

\end{thebibliography}
\end{document}